\documentclass[preprint,3p,12pt]{elsarticle}

\usepackage{amssymb}
\usepackage{amsmath}
\usepackage{float} 
\usepackage{subcaption} 
\usepackage{hyperref}
\usepackage{xurl}

\makeatletter
\def\ps@pprintTitle{%
    \let\@oddfoot\@empty
    \let\@evenfoot\@empty
}
\makeatother

\begin{document}


\begin{frontmatter}


\title{A modeling framework to support the electrification of private transport in African cities: a case study of Addis Ababa}

\author[add1]{\texorpdfstring{Jérémy Dumoulin\corref{cor1}}{Jérémy Dumoulin}}
\ead{jeremy.dumoulin@epfl.ch}
\author[add2]{Dawit Gebremeskel}
\author[add3]{Kanchwodia Gashaw}
\author[add4]{Ingeborg Graabak}
\author[add1]{Noémie Jeannin}
\author[add1]{Alejandro Pena-Bello}
\author[add1]{Christophe Ballif}
\author[add1]{Nicolas Wyrsch}

\cortext[cor1]{Corresponding author.}
\address[add1]{Photovoltaics and thin film electronics laboratory (PV-LAB), École Polytechnique Fédérale de Lausanne (EPFL), Institute of Electrical and Microengineering (IEM), Neuchâtel, Switzerland}
\address[add2]{School of Electrical and Computer Engineering, Addis Ababa Institute of Technology, Addis Ababa University, Addis Ababa, Ethiopia}
\address[add3]{Veritas Consulting, Addis Ababa, Ethiopia}
\address[add4]{SINTEF Energy Research, Trondheim, Norway}

\begin{abstract}
The electrification of road transport, as the predominant mode of transportation in Africa, represents a great opportunity to reduce greenhouse gas emissions and dependence on costly fuel imports. However, it introduces major challenges for local energy infrastructures, including the deployment of charging stations and the impact on often fragile electricity grids. Despite its importance, research on electric mobility planning in Africa remains limited, while existing planning tools rely on detailed local mobility data that is often unavailable, especially for privately owned passenger vehicles. In this study, we introduce a novel framework designed to support private vehicle electrification in data-scarce regions and apply it to Addis Ababa, simulating the mobility patterns and charging needs of 100,000~electric vehicles. Our analysis indicate that these vehicles generate a daily charging demand of approximately 350~MWh and emphasize the significant influence of the charging location on the spatial and temporal distribution of this demand. Notably, charging at public places can help smooth the charging demand throughout the day, mitigating peak charging loads on the electricity grid. We also estimate charging station requirements, finding that workplace charging requires approximately one charging point per three electric vehicles, while public charging requires only one per thirty. Finally, we demonstrate that photovoltaic energy can cover a substantial share of the charging needs, emphasizing the potential for renewable energy integration. This study lays the groundwork for electric mobility planning in Addis Ababa while offering a transferable framework for other African cities.
\end{abstract}

\end{frontmatter}



\section{Introduction} 
\label{introduction}
Africa's road-dependent transport sector is rapidly expanding \cite{unep_green_growth_transport, DecarbAfrica}, with demand for transport fuels projected to increase by two-thirds by 2040 under current policies and trajectories \cite{unep_green_growth_transport}. This growing reliance on fossil fuels presents critical economic and environmental challenges, particularly as many African nations strive to meet their climate commitments under the Paris Agreement. Several countries are setting targets and policies to promote low-emission mobility solutions. Ethiopia, for example, as part of its Climate Resilient Green Economy (CRGE) strategy, aims to develop a greener transport sector by encouraging fuel-efficient and sustainable transport solutions \cite{fdre_crge_2011}. In this context, electric vehicles (EVs) emerge as a promising solution, leveraging Africa's vast renewable energy potential. EV adoption could also provide economic benefits by reducing dependence on costly fossil fuel imports and the associated forex requirement. In Ethiopia alone, fossil fuel imports amounted to nearly 4~billion~USD in 2023, accounting for 23\% of total imports \cite{undp_2024}, most of which is consumed by the transport sector \cite{IEA_Ethiopia_2024}. To mitigate the country's recent forex shortages, the Ethiopian government has implemented a progressive lift of fuel subsidies over the last two years, particularly on private vehicles. Ethiopia's National Bank has reported an 18.3 percent increase in the average retail price of fuel in 2024, compared to the previous year, resulting from the lifting of fuel subsidies as well as global fluctuations in the price of fuel. The transition to electric mobility could, therefore, play a key role in advancing both climate and economic sustainability across the continent.

\begin{figure}[htbp]
    \centering
    \includegraphics[width=0.6\textwidth]{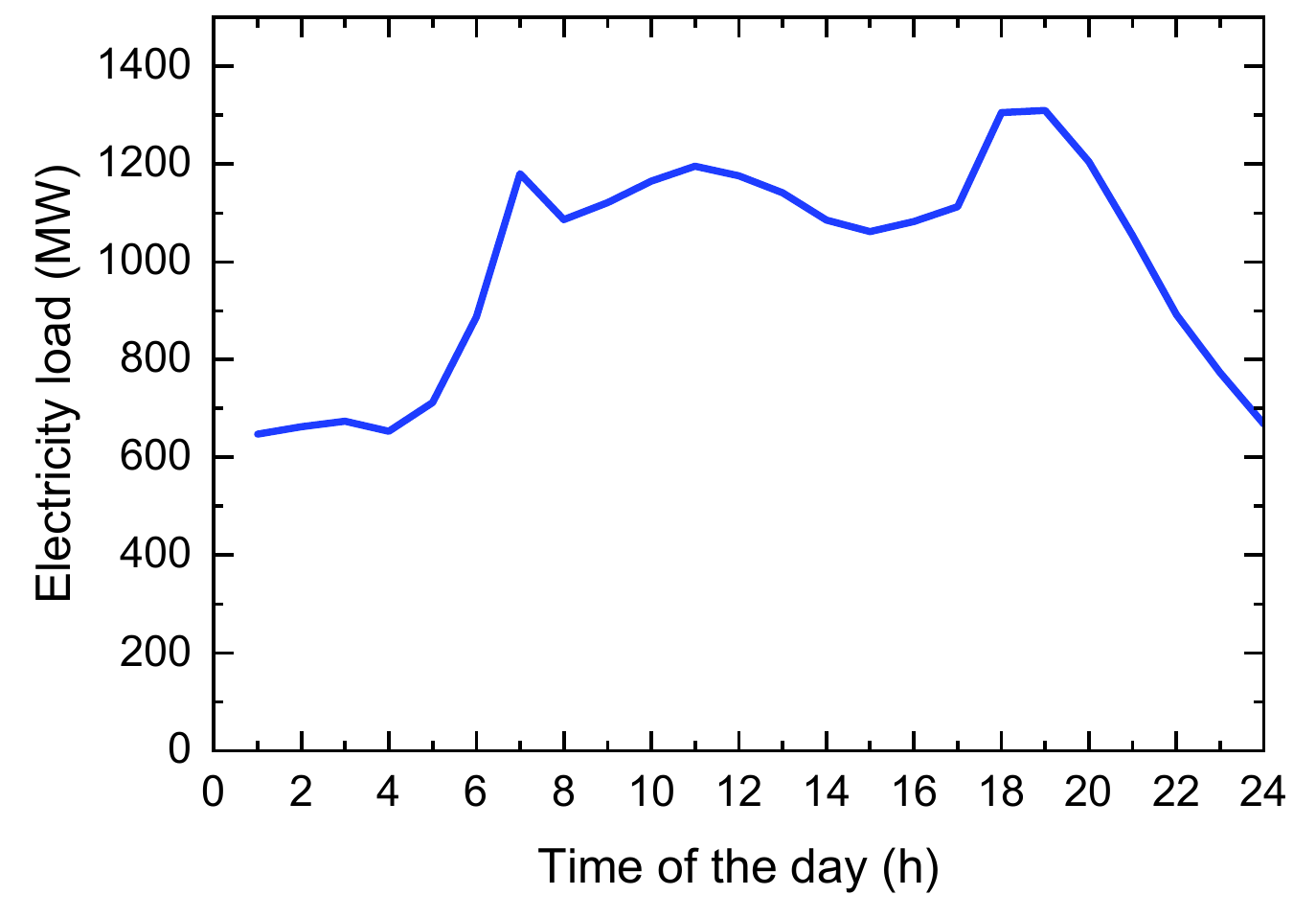}            
    \caption{Hourly electricity demand profile for Addis Ababa on a typical weekday, derived from national demand data provided by the National Grid Control Center and scaled to the city level. The scaling methodology is detailed in \protect\ref{app:demand_profile}.}
    \label{fig:aa_load_profile}
\end{figure}

Against this background, Addis Ababa - the capital of Ethiopia - emerges as a key city for studying the electrification of privately owned vehicles. Not only does it stand out for its political and economic importance, but also for symbolizing a dual narrative with transformative opportunities as well as critical challenges. On the one hand, Ethiopia has taken a pioneering role in becoming the first country to implement a total ban on fossil fuel car imports. In line with this ambitious policy, the government has introduced financial incentives to promote electric vehicles, including import tax reductions, exemptions from VAT, excise, and surtaxes \cite{Eticha2023, etmp_2022, efficient_vehicles_policy_2021}. Over the next decade, the country aims to import several hundred thousand EVs \cite{efficient_vehicles_policy_2021}. Given that Addis Ababa is home to more than half of Ethiopia's 1.2~million vehicles \cite{Eticha2023}, it is expected to absorb the majority of these imports. On the other hand, there are significant concerns about whether local energy infrastructures can support the large-scale adoption of EVs. Currently, Addis Ababa has only a handful of public charging stations \cite{electromaps_2024}, and the rapid expansion of home charging facilities may further strain Addis Ababa's power system, which already faces frequent electricity supply interruptions \cite{Tsegai2022, Meles2020}. Figure \ref{fig:aa_load_profile} presents the hourly electricity demand curve of Addis Ababa. As shown in the figure, the grid undergoes high electricity demand fluctuations. During peak periods, it becomes challenging for the grid to match supply and demand, increasing the risk of power instability and interruptions. Studies are therefore urgently needed to guide policies and strategic infrastructure planning, ensuring that Ethiopia's ambitious electrification efforts translate into a smooth and sustainable transition.

Effective planning for the electrification of private transport requires tailored approaches that account for the unique energy and mobility landscapes of different regions. However, there is a notable scarcity of studies on electric mobility in Africa, particularly regarding the electrification of private vehicles. Most of the recent work is centered on the electrification of informal public transport systems \cite{Collett2021, Rix2022, Giliomee2023}, leaving the adoption and implications of private EV usage underexplored. Yet, even a modest increase in private EV adoption has the potential to impose considerable stress on local power grids \cite{Mudaheranwa2023}. Existing research also lacks comprehensive methodologies capable of providing actionable, region-specific insights for policymakers or planners. While some studies have investigated strategies like vehicle-to-grid (V2G) solutions \cite{McPherson2018, Bouguerra2019} or coordinated EV charging to mitigate peak electricity demand \cite{Dioha2022b, Makeen2023} in Africa, these analyses do not consider the spatial and temporal distribution of the charging demand. Furthermore, they rarely explore the integration of local renewable energy sources, such as photovoltaic (PV) systems, which could play an important role in meeting the additional electricity demand. The absence of practical recommendations regarding the required charging infrastructure also further underscores the need for studies that go beyond only quantifying the EV charging needs. Additionally, most existing studies are limited in their applicability to regions with scarce mobility data, such as Addis Ababa. State-of-the-art methodologies for assessing EV charging needs mainly rely on household travel surveys \cite{Pareschi2020, Liu2022}, data that is unavailable in many developing countries. In the case of Addis Ababa, previous research has only focused on driving profiles \cite{Mamo2023} or agent-based simulations to estimate energy consumption and CO$_2$ emission reductions \cite{Eticha2023}, without offering a general methodology to support private vehicle electrification.

In this study, we propose a novel modeling framework specifically designed for regions with limited mobility data, demonstrating its application in the case of Addis Ababa. The framework leverages open-source geospatial data to estimate EV charging needs and their spatio-temporal distribution. Based on these estimates, the implications for charging infrastructure requirements and the potential contribution of PV energy in meeting the charging needs are also quantified. Specifically, we use our model to assess the following three research questions:
\begin{itemize}
    \item What is the additional electricity demand from private EV charging, and how is it distributed spatially and temporally, considering different charging locations?
    \item What are the implications for Addis Ababa in terms of charging infrastructure requirements? 
    \item How can locally installed PV systems contribute to meeting this additional demand?
\end{itemize}

The paper is structured as follows. Section 2 introduces the proposed modeling framework for electric vehicle planning. Section 3 demonstrates its application to Addis Ababa, providing insights into the local context and the crucial factors for EV planning studies. Section 4 presents the results and discussion, while Section 5 concludes the study and highlights directions for future research. This work not only provides critical insights into the infrastructure planning needs for Addis Ababa but also offers a transferable methodology for electric mobility planning in other cities.

\section{Methodology} 
\label{methodology}

\begin{figure}[ht]
    \centering
    \includegraphics[width=1.0\textwidth]{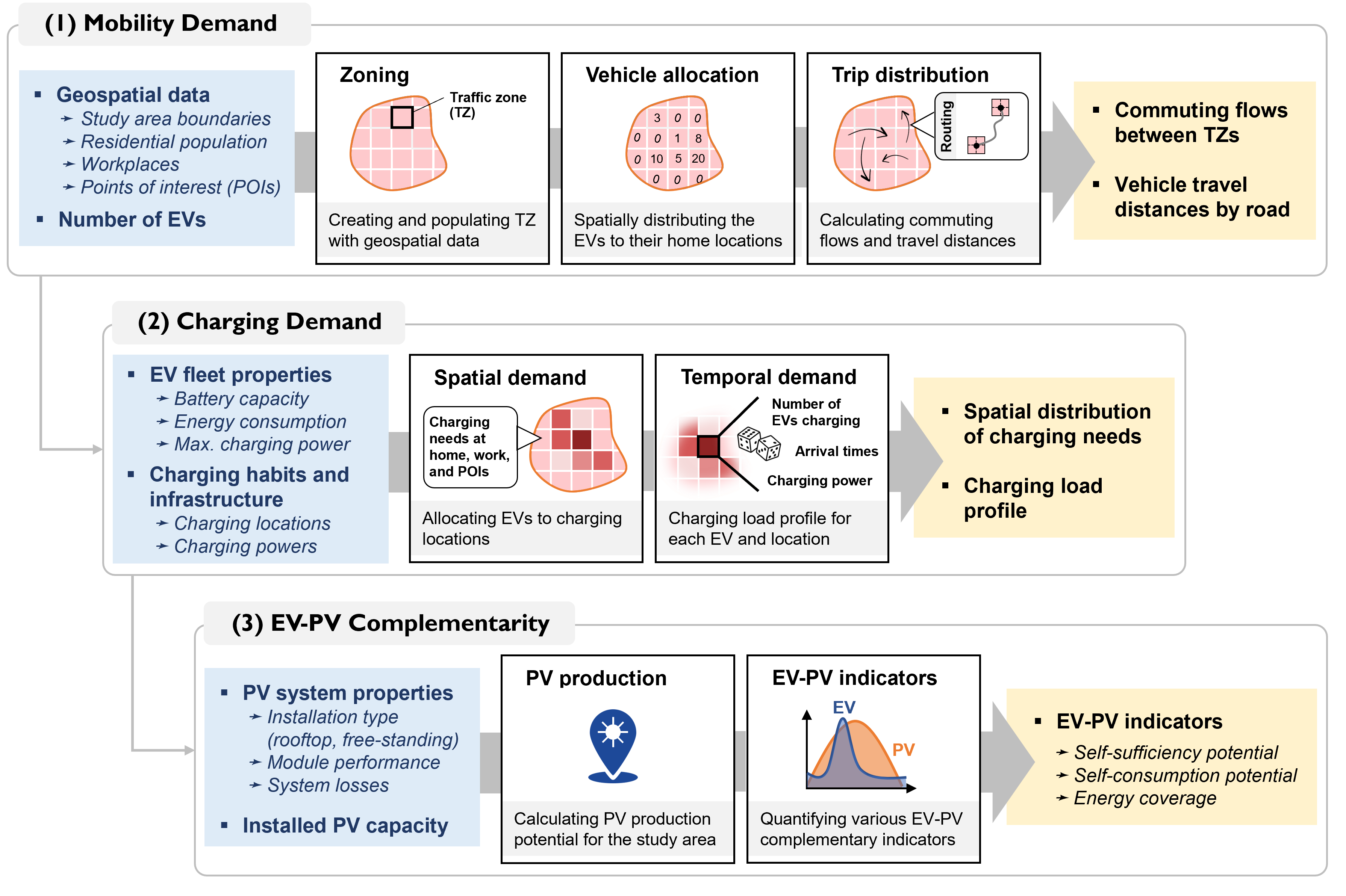}            
    \caption{Overview of the three main steps of the modeling framework}
    \label{fig:workflow_schematic}
\end{figure}

The proposed model provides a comprehensive and integrated framework to analyze passenger mobility demand, EV charging needs, the potential of PV energy to meet these needs, and the required charging infrastructure across various charging locations. By providing detailed modeling of intra-urban mobility patterns, the model is particularly well-suited for supporting electric mobility planning in cities. As illustrated in Fig.~\ref{fig:workflow_schematic}, the framework is structured around three main steps:
\begin{enumerate}
    \item Mobility demand: This step simulates daily mobility patterns with a specific focus on detailed spatial modeling of home-to-work commuting flows through the use of geospatial data.
    \item Charging demand: The model then determines the spatio-temporal distribution of EV charging needs and estimates the required number of charging points. It relies on EV fleet properties, defined for each vehicle type (battery capacity, energy consumption per kilometer, and maximum charging power), as well as user charging habits (share of EVs charging at home, work and public places) and charging infrastructure properties (available charging power at each location).
    \item EV-PV complementarity: Finally, the complementarity between local PV production and EVs is assessed. Using basic parameters of local PV installations, the model evaluates how effectively PV energy can fulfill EV charging needs.
\end{enumerate}

Designed to be flexible and operable with open-source geospatial data, the framework is particularly well suited for regions with limited data availability or uncertain development trajectories, making it a versatile tool for strategic decision-making in various contexts. The following sections provide a detailed description of each modeling step.

\subsection{Mobility demand}

The first step involves quantifying the daily mobility demand of electric vehicles using a spatially-dependent mobility modeling approach. This process involves dividing the study area into smaller, manageable units, referred to as traffic zones (TZs). A specific number of EVs is allocated to each TZ, and a trip distribution model is applied to simulate vehicle flows and travel distances between these zones. 

The model primarily focuses on capturing weekday mobility patterns, with a particular emphasis on home-to-work commuting and the associated charging needs at home, work, and public charging locations. Home-to-work trips are expected to dominate the weekday mobility demand \cite{eurostat_passenger_2021}, justifying their central role in the modeling framework. However, to ensure applicability across various case studies, the model also allows for the incorporation of additional travel distances for other trip purposes, based on external inputs.

\subsubsection{Zoning}

The zoning process involves dividing the study area into TZs, which serve as the base geographical units for the model. A regular zoning scheme is adopted \cite{Ghadiri2019}, where each traffic zone is of equal size and shape, ensuring simplicity and uniform spatial granularity. This zoning scheme also facilitates seamless integration with the self-calibrating trip distribution model introduced in section 2.1.3. Subsequently, the traffic zones are populated with the relevant geospatial data, aggregated at the zone level: residential population, number of workplaces, and number of points of interest (POIs).

\subsubsection{Vehicle allocation}

Following the zoning process, the next step involves populating the different zones with the number of EVs to be simulated. This vehicle allocation step is based on residential population, assuming that the number of vehicles $n_i$ of a traffic zone $i$ is proportional to the number of people $P_i$ of that zone
\begin{equation}
n_i = \frac{P_i}{P_{tot}} \cdot n_{tot} 
\end{equation}
with $n_{tot}$ the total number of simulated EVs. The resulting number of EVs per zone, $n_i$, is rounded to the nearest integer, while ensuring that the total allocated vehicles matches the target number $n_{tot}$.

\subsubsection{Trip distribution}
The vehicle flows for home-to-work trips between TZs are estimated using a production-constrained spatial interaction model \cite{Fotheringham2001}, based on a gravity law with a decaying exponential function. Given the number of vehicles commuting from a specific traffic zone (origin), the model calculates the number of trips directed towards other zones (destinations) by assuming that the number of trips between two zones is proportional to the so-called attractiveness of the destination (equal to the number of workplaces) and decays with distance, following an exponential law. This approach was selected because it has been shown to outperform other models for trip distribution, particularly at small spatial scales \cite{Liang2013, Lenormand2016}. Numerically, the trip distribution model calculates the probability of a vehicle traveling from zone $i$ to zone $j$
\begin{equation}
    p_{ij}=c_i \cdot A_j \cdot e^{-\beta d_{ij}}
\end{equation}
where $A_j$ is the attractiveness of zone $j$, $d_{ij}$ is the inter-zone distance, $\beta$ is a free parameter, and $c_i$ is a normalization constant ensuring that the probabilities for all possible destinations sum to 1. 

Importantly, the parameter $\beta$ is determined using a self-calibrating approach, which eliminates the need for manual calibration of the trip distribution model, an essential feature for regions where detailed mobility data is unavailable. Based on the extensive work of Lenormand et al. \cite{Lenormand2016}, $\beta$ is defined as:
\begin{equation}
    \beta=0.3 \cdot S^{-0.18}
\end{equation}
where $S$ represents the surface area of the traffic zones in km$^2$. 

Road distances between TZs are calculated using Open Route Service \cite{openrouteservice}, with traffic zone centroids as reference points for the routing algorithm. When routing between two traffic zones is unavailable, a circuity factor is used to estimate the inter-zonal distance. The circuity factor \cite{Ballou2002}, defined as the ratio of road distance to Euclidean distance, is derived by performing routing between a set of random points within the study area and applying linear regression to calculate the average factor.

\subsection{Charging demand}

The spatial and temporal estimation of the EV charging demand is performed in two modeling steps, each using a distinct methodology. First, the daily charging demand is calculated for each traffic zone and charging location category (home, work, and POIs). Second, a stochastic modeling approach is used to determine the temporal charging demand, accounting for the variability in the EV charging load profiles.

\subsubsection{Spatial demand}
The spatial charging demand for a given traffic zone $i$ is segmented by charging location: at home ($E_{home,i}$), at work ($E_{work,i}$), and at at POIs ($E_{POI,i})$. The total demand is the sum of these components.

For home and work, the demand is calculated as the product of the vehicle kilometers ($V\!K\!M$) derived from the mobility demand model, the corresponding charging location shares ($f_{home}$ and $f_{work}$), the average energy consumption per kilometer of EVs ($C_{EV}$), and the inverse of the charging efficiency ($\eta_{charge}$):
\begin{equation}
\label{eq:spatial_home}
    E_{home,i}=f_{home} \cdot C_{EV} \cdot \eta_{charge}^{-1} \cdot V\!K\!M_{out,i}
\end{equation}
\begin{equation}
\label{eq:spatial_work}
    E_{work,i}=f_{work} \cdot C_{EV} \cdot \eta_{charge}^{-1} \cdot V\!K\!M_{in,i}
\end{equation}
Here, $V\!K\!M_{out,i}$  represents the distance traveled by vehicles from traffic zone $i$ to other zones, while $V\!K\!M_{in,i}$ refers to the distance traveled by vehicles from other traffic zones toward zone $i$ as a destination.

For POIs, the estimation of charging needs follows a similar logic. However, since the mobility model does not specify how many vehicles stop at each POI, the charging demand is distributed proportionally to the number of POIs in each TZ. This approach ensures that zones with more POIs are allocated a larger share of the total POI-related charging demand. The distribution is given by:
\begin{equation}
    E_{POI,i} = \left( f_{POI} \cdot C_{EV} \cdot \eta_{charge}^{-1} \cdot \sum_i V\!K\!M_{out,i} \right) \cdot \frac{m_i}{m_{tot}},
\end{equation}
where $f_{POI}$ denotes the share of EVs charging at POIs, $m_i$ is the number of POIs in zone $i$, and $m_{tot}$ is the total number of POIs across all zones.

In equations \ref{eq:spatial_home} and \ref{eq:spatial_work}, the $V\!K\!M$ is derived from the mobility demand model as follows:
\begin{equation}
    V\!K\!M_{out,i} = 2 \cdot \sum_{j \neq i} p_{ij} \cdot n_{i} \cdot d_{ij},
\end{equation}
\begin{equation}
    V\!K\!M_{in,i} = 2 \cdot \sum_{j \neq i} p_{ji} \cdot n_{j} \cdot d_{ij},
\end{equation}
where $n_{i}$ is the number of EVs of traffic zone $i$, and $d_{ij}$ is the distance between zones $i$ and $j$. The factor of 2 accounts for round trips (outward and return trip to work). 

\subsubsection{Temporal demand}
The temporal charging demand (i.e., the charging load profile) is determined by estimating the number of EVs charging on a given day and constructing the charging load profile for each individual vehicle. Both processes employ a stochastic modeling approach, ensuring that the charging load profile reflects behavioral heterogeneity and day-to-day variability. In this work, we assume an uncontrolled charging scheme, where vehicles begin charging immediately upon arrival at a constant power rate. However, the model is designed to be adaptable to controlled charging schemes, which will be explored in future investigations.

To estimate the number of EVs charging per day, we build on the experimentally validated model introduced by Pareschi et al. \cite{Pareschi2020}, which introduces a threshold state of charge ($SoC_0$), below which vehicles are assumed to initiate charging. For each EV, this threshold is randomly assigned based on Pareschi's probabilistic distribution, which follows a normal distribution with a mean of 0.6 and a standard deviation of 0.2. Then, an average number of days between charging events ($\Delta N$) is computed based on the daily state-of-charge depletion ($\Delta SoC_{daily}$) of each EV
\begin{equation}
\Delta N = \max \left( 1, \frac{1-SoC_0}{\Delta SoC_{daily}} \right)
\end{equation}
\begin{equation}
\Delta SoC_{daily} = \frac{E_{daily}}{0.8 \cdot Q}
\end{equation}
Here, $E_{daily}$ denotes the vehicle's daily energy demand, assigned randomly from the daily travel distance distribution of the TZ the vehicle belongs to, while $Q$ represents the nominal battery capacity. The factor 0.8 accounts for the useful battery capacity for decision-making, in line with Pareschi's empirically calibrated model \cite{Pareschi2020}. 

It is important to note that our approach slightly differs from the original model, as we directly assume $\Delta N$ for each EV, whereas Pareschi's work does not assume such a steady-state condition a priori. This assumption eliminates the need for multiple preliminary runs to reach model convergence. Once $\Delta N$ is determined, the model assigns a daily charging probability of $1 / \Delta N$ to each vehicle and a random sampling process identifies whether a vehicle charges on a given day. This approach moves beyond simplistic assumptions (e.g., daily charging) by accommodating the behavior of vehicles with large battery capacities or low daily travel distances. For illustration, a plot of the daily charging probability as a function of the daily EV energy use for two battery capacities $Q$ is provided in \ref{app:charging_probability}.

For vehicles identified as charging on a given day, a 24-hour charging load profile is constructed by sampling several parameters based on the modeled scenario. Specifically, the charging location is assigned according to the distribution of home, work, and POIs charging shares. The arrival time ($t_{arrival}$) is sampled from the probability distribution of arrival times specific to the assigned charging location. The charging power ($P_{charge}$) is determined by the availability of chargers at the location and is capped by the vehicle's maximum charging power ($P_{max}$). Using these parameters, the charging load profile for a vehicle $v$, $P_{EV,v}(t)$, is defined as follows:
\begin{equation}
P_{EV,v}(t) = 
\begin{cases} 
    P_{\text{charge}} & \text{if } t \in [t_{\text{arrival}}, t_{\text{end}}] \\
    0 & \text{otherwise}
\end{cases}
\end{equation}
\begin{equation}
t_{end} = t_{arrival} + \frac{\Delta N \cdot E_{daily}}{\eta_{charge} \cdot P_{charge}}
\end{equation}
Here, $t_{end}$ denotes the end time of the charging session, which is calculated based on the charging duration while accounting for the charging efficiency $\eta_{charge}$.

\subsection{EV-PV complementarity}
The final step in the analysis evaluates the potential complementarity between PV energy and EVs, focusing primarily on the ability of locally installed PV systems to meet EV charging needs. This section comprises two components: the computation of time-dependent local PV production and the calculation of key indicators that quantify EV-PV complementarity.

\subsubsection{PV production potential}
The local PV production potential is calculated using the pvlib toolbox \cite{Anderson2023}, which enables detailed yet computationally efficient simulations by integrating environmental conditions, PV module specifications, and installation parameters. For environmental conditions, we use local weather data from the PVGIS-SARAH3 solar database, which can be directly extracted through pvlib \cite{Jensen2023}.

For each time step, the plane-of-array irradiance ($G_{\text{POA}}(t)$) is calculated based on irradiance data and the PV module's installation angles (tilt and azimuth). These angles can be customized for fixed-tilt or tracking systems. The resulting irradiance is then used to determine the PV production $P_{\text{PV}}(t)$, calculated using the so-called \textit{PVWatts} model from pvlib to capture thermal losses, as shown hereafter
\begin{equation}
P_{\text{PV}}(t) = \eta_{\text{PV}} \cdot G_{\text{POA}}(t) \cdot \left[ 1 + \beta \cdot \left( T_{\text{cell}}(t) - T_{\text{ref}} \right) \right]
\end{equation}
with $\eta_{PV}$ the nominal module efficiency, $\beta$ the temperature coefficient, $T_{cell}$ the solar cells operating temperature (calculated using the pre-implemented \textit{PVsyst} model), and $T_{ref}$ the reference temperature equal to 25 $^{\circ}$C.

Angular losses are also automatically accounted for at each time step using the analytical model of Martin and Ruiz \cite{Martin2001}. Additional system losses are added ex-post by an overall loss factor. This methodology includes all main loss mechanisms ensuring a robust assessment of the local PV production potential under real-world conditions.

\subsubsection{EV-PV indicators}
After estimating PV generation and EV charging demand, key indicators are used to assess their complementarity. These include self-sufficiency potential, self-consumption potential, and energy coverage (ratio of the total PV production to the EV charging demand over a given period).

In this work, we focus on the self-sufficiency potential ($SS$), which quantifies the proportion of the charging demand that can be met by PV production, assuming all available PV energy can be used for charging. It is calculated as the ratio of the coincident power (i.e., the overlap between PV generation and EV demand) to the total EV charging demand over a 24-hour period:
\begin{equation} 
    SS = \frac{\int \min[ P_{PV}(t), P_{EV}(t) ] dt}{\int P_{EV}(t) dt} 
\end{equation}
where $P_{EV}(t)$ is the aggregated charging demand for all EVs. This metric treats all PV generation and EV demand as part of a closed system, capturing the overall balance between PV production and EV charging demand.


\section{Case study} 
\label{casestudy}

\subsection{Study area}
\begin{figure}[htbp]
    \centering
    \includegraphics[width=0.4\textwidth]{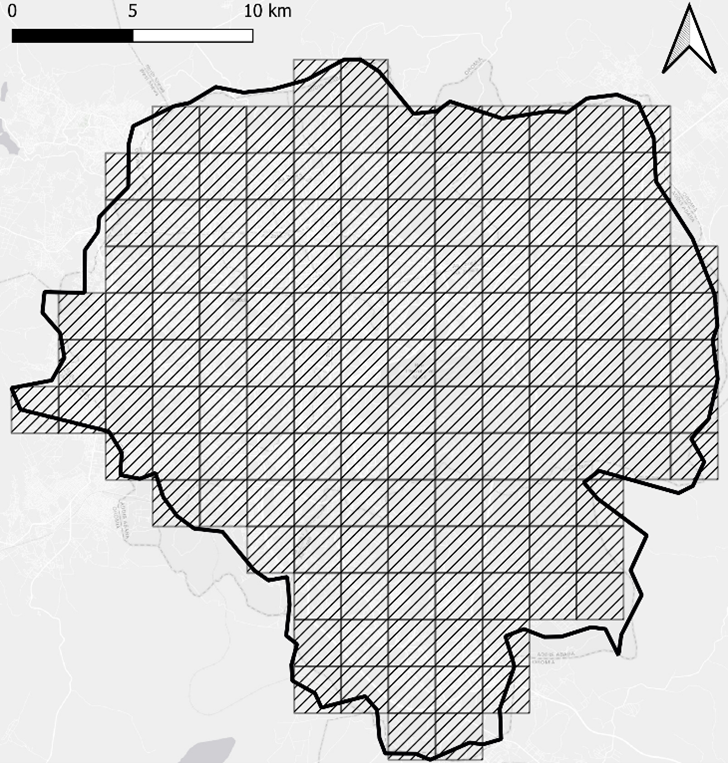}            
    \caption{Map of the study area (Addis Ababa) showing administrative boundaries and traffic zones.}
    \label{fig:addis_ababa_studyarea}
\end{figure}

Figure \ref{fig:addis_ababa_studyarea} illustrates the study area, defined by the administrative boundaries of Addis Ababa, which cover approximately 540~km$^2$ \cite{WUBNEH2013}. These boundaries were obtained from the Global Administrative Areas (GADM) dataset \cite{gadm}. The area is divided into traffic zones measuring 1.95~km by 1.95~km, a grid resolution selected to effectively capture short-distance commuting patterns while ensuring manageable computational time.

Geospatial data, aggregated at the zone level, includes residential population figures from the GHS-POP dataset \cite{ghs-pop}, along with the number of workplaces and points of interest obtained from OpenStreetMap \cite{openstreetmap}. The workplaces and POIs are carefully extracted to represent the majority of potential commuting destinations and charging locations (see the \ref{app:osm}). Notably, universities are also included as key destinations, reflecting their role in accommodating a non-negligible share of Addis Ababa's commuting population. The resulting dataset reveals 1845 workplaces, and 3633 POIs, alongside with a residential population of 5.54~million people.

\subsection{Electric vehicle fleet}
The simulated EV fleet consists of 100,000~vehicles, representing a realistic short-term projection in light of Ethiopia's expanding private vehicle market and the government's national target. As shown in \ref{app:ev_fleet_share}, even under conservative assumptions regarding the current EV stock, car renewal rates, and car ownership growth, this target could be achieved within approximately 3.6~years if all newly registered vehicles are EVs. The technical characteristics of the simulated EVs are summarized in Table \ref{tab:ev_parameters}. The fleet composition includes 80\% battery electric vehicles (BEVs) and 20\% plug-in hybrid electric vehicles (PHEVs), closely aligning with the distribution observed in African EV sales \cite{EVOutlook2024}. Incorporating a share of PHEVs also reflects a realistic assumption regarding range anxiety faced by drivers due to Addis Ababa's emerging and limited charging infrastructure. Due to the lack of statistical data, battery capacities were selected based on informed assumptions, consistent with those used in the latest International Energy Agency report \cite{EVOutlook2024}. Additionally, PHEVs are limited to a maximum charging power of 11~kW to restrict their access to fast-charging stations.

\begin{table}[ht]
\centering
\setlength{\tabcolsep}{14pt} 
\caption{Parameters of the simulated 100,000 EVs.}
\begin{tabular}{llll}
\hline
Parameter & BEV & PHEV & Reference \\ \hline \hline
Share (\%) & 80 & 20 & \cite{EVOutlook2024} \\ 
Battery capacity (kWh) & 60 & 15 & \cite{EVOutlook2024} \\ 
Energy consumption (kWh/km) & 0.183 & 0.183 & \cite{Fetene2017} \\ 
Maximum charging power (kW) & None & 11 & Own assumption \\ \hline
\end{tabular}
\label{tab:ev_parameters}
\end{table}

\subsection{Charging habits and infrastructure}
\subsubsection{Charging scenarios}
In Addis Ababa, the future evolution of charging habits and the availability of charging infrastructure remain highly uncertain. In this context, a key question involves understanding how different charging locations might influence the spatial and temporal distribution of EV charging. This, in turn, affects the potential grid impacts and the charging infrastructure requirements, which are crucial factors for informing policy decisions and energy infrastructure planning. To explore this question, we analyze three archetypal charging scenarios, each representing distinct charging location patterns:
\begin{itemize}
    \item \textbf{100\% home:} This scenario assumes that all users charge their vehicles at home. It is typical for areas with accessible residential infrastructure and favorable overnight tariffs, reflecting standard overnight charging demand patterns.
    \item \textbf{100\% work:} In this scenario, all charging takes place at work, reflecting the behavior of users who do not have access to home charging and rely entirely on employer-provided charging stations. 
    \item \textbf{Mixed:} Charging is distributed across three locations (25\% at home, 25\% at workplaces, and 50\% at public POIs such as shopping centers and leisure facilities). This scenario captures a more diverse set of charging behaviors and highlights the impact of a reliance on the public charging infrastructure.
\end{itemize}

Arrival times at the various charging locations are modeled using a normal distribution centered around typical arrival times. For home charging, the mean arrival time is assumed to be 6:00~PM, with a standard deviation of 2.7~hours. For work charging, the mean arrival time is set at 9:00~AM, with a standard deviation of 1.8~hours. In the absence of local data, these average arrival times are based on informed estimates from the authors who reside in Addis Ababa, while the standard deviations are derived from Swiss microcensus data \cite{microcensus2021}. For charging at POIs, we assume that charging occurs randomly between the arrival time at work and the return home.

\subsubsection{Charging infrastructure}
Each charging location incorporates a diverse mix of power levels, informed by established proxy data (Table \ref{tab:charger_availability}). Home charging primarily consists of slow (3.2~kW) and moderate (7.4~kW and 11~kW) chargers, following the distribution reported by Sorensen et al. \cite{Sorensen2023}. POI charging is based on Swiss data for public charging infrastructure \cite{je_recharge_mon_auto}. Notably, these charging power distributions align closely with data from other countries \cite{lucas2024, NREL_EVtrends}, supporting the validity of the assumed charging power mix for this case study. Workplace charging is modeled with only Level 2 chargers \cite{NREL_EVtrends}, with 11~kW units expected to dominate.

\begin{table}[ht]
\centering
\setlength{\tabcolsep}{14pt} 
\caption{Charger option availability at different locations.}
\begin{tabular}{lccc}
\hline
Charger Option & Home & Work & POIs \\ \hline \hline
3.2 kW & 45\% & - & - \\ 
7.4 kW & 40\% & 25\% & 15\% \\ 
11 kW & 15\% & 50\% & 15\% \\ 
22 kW & - & 25\% & 55\% \\ 
50 kW & - & - & 15\% \\ \hline
\end{tabular}
\label{tab:charger_availability}
\end{table}

Following the approach of Lanz et al. \cite{Lanz2022}, fast charging is capped at 50~kW due to the limited availability of higher-power charging stations and the observation that, even for higher charging power levels, the average charging power typically remains below the nominal values. Additionally, a uniform charging efficiency of 90\% is applied across all locations \cite{Sears2014}.

\subsection{PV system}
This study considers free-standing PV systems configured with an optimal tilt and azimuth angle, representative of typical installations in PV farms or standalone systems for solar-powered charging stations. Weather data for Addis Ababa was obtained for the year 2020. The PV system is modeled with a nominal module efficiency of 22\%, reflecting the recent sales-weighted average for commercially available PV modules \cite{FraunhoferISE_PVReport}. To account for real-world performance losses, a temperature coefficient of -0.4\%/$^{\circ}$C \cite{Jan2017} and typical system losses of 14\% were incorporated into the simulation. As a result, the model predicts an annual energy yield of 1656.4~kWh/kWp, with the lowest energy generation observed during the rainy season, which occurs from July to September. 


\section{Results \& discussion} 
\label{results}

\subsection{Spatio-temporal charging demand}

\subsubsection{Spatial distribution}
\begin{figure}[htbp!]
    \centering
    \begin{minipage}[b]{1.0\textwidth} 
        \centering
        \begin{subfigure}[t]{0.42\textwidth}
            \centering
            \caption{}
            \includegraphics[width=\textwidth]{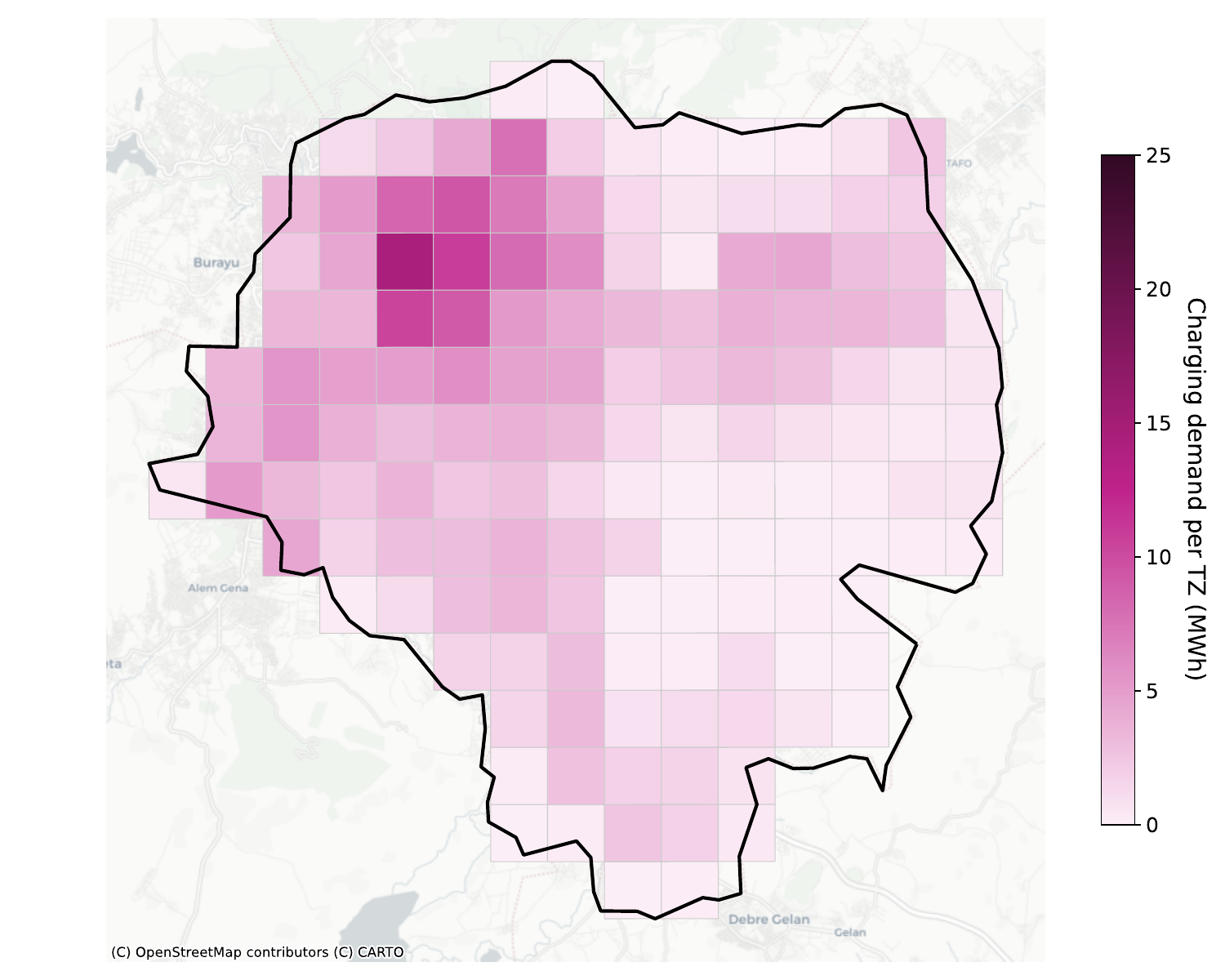}            
        \end{subfigure}
        \begin{subfigure}[t]{0.42\textwidth}
            \centering
            \caption{}
            \includegraphics[width=\textwidth]{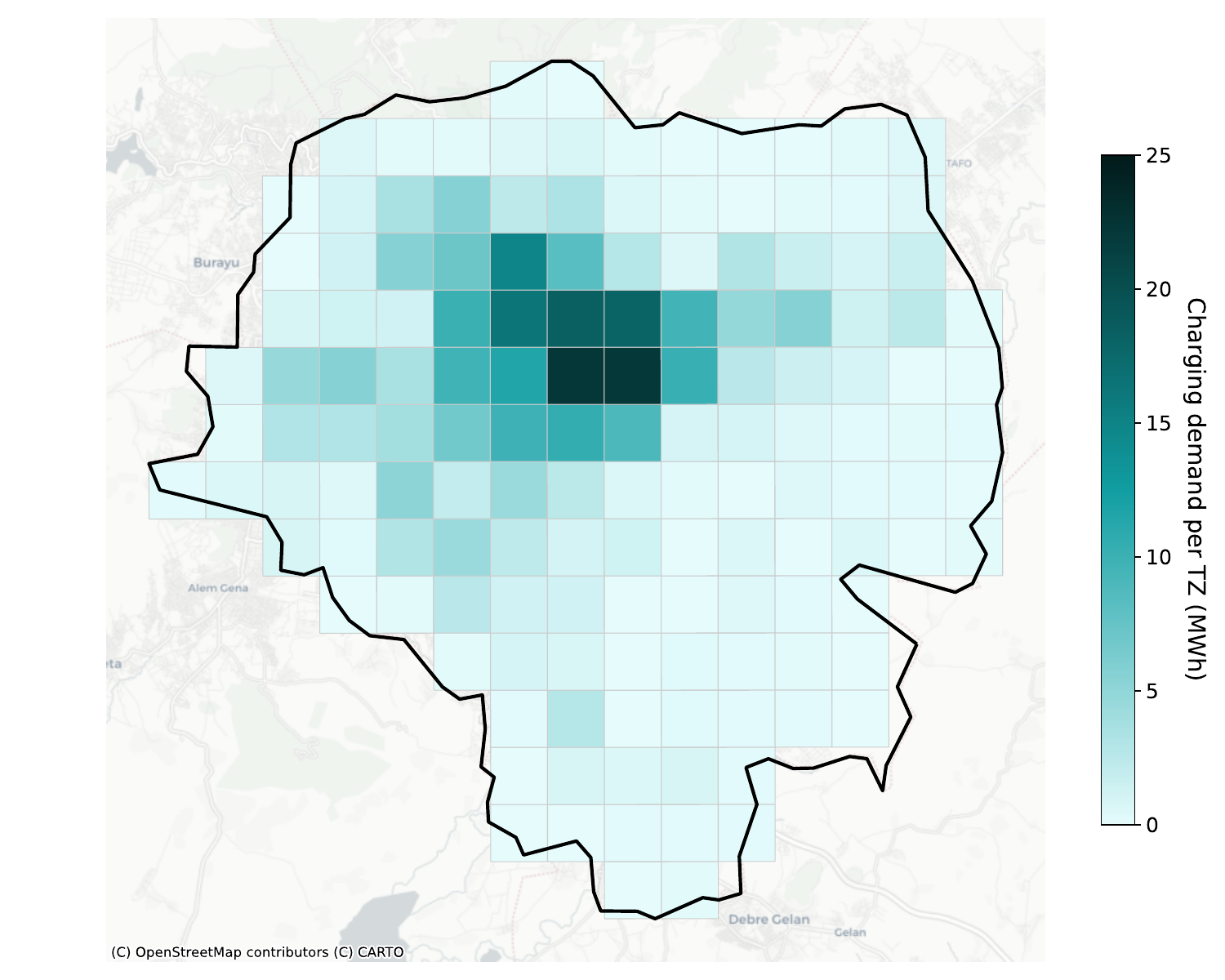}            
        \end{subfigure}
        \hfill
        \begin{subfigure}[t]{0.42\textwidth}
            \centering
            \caption{}
            \includegraphics[width=\textwidth]{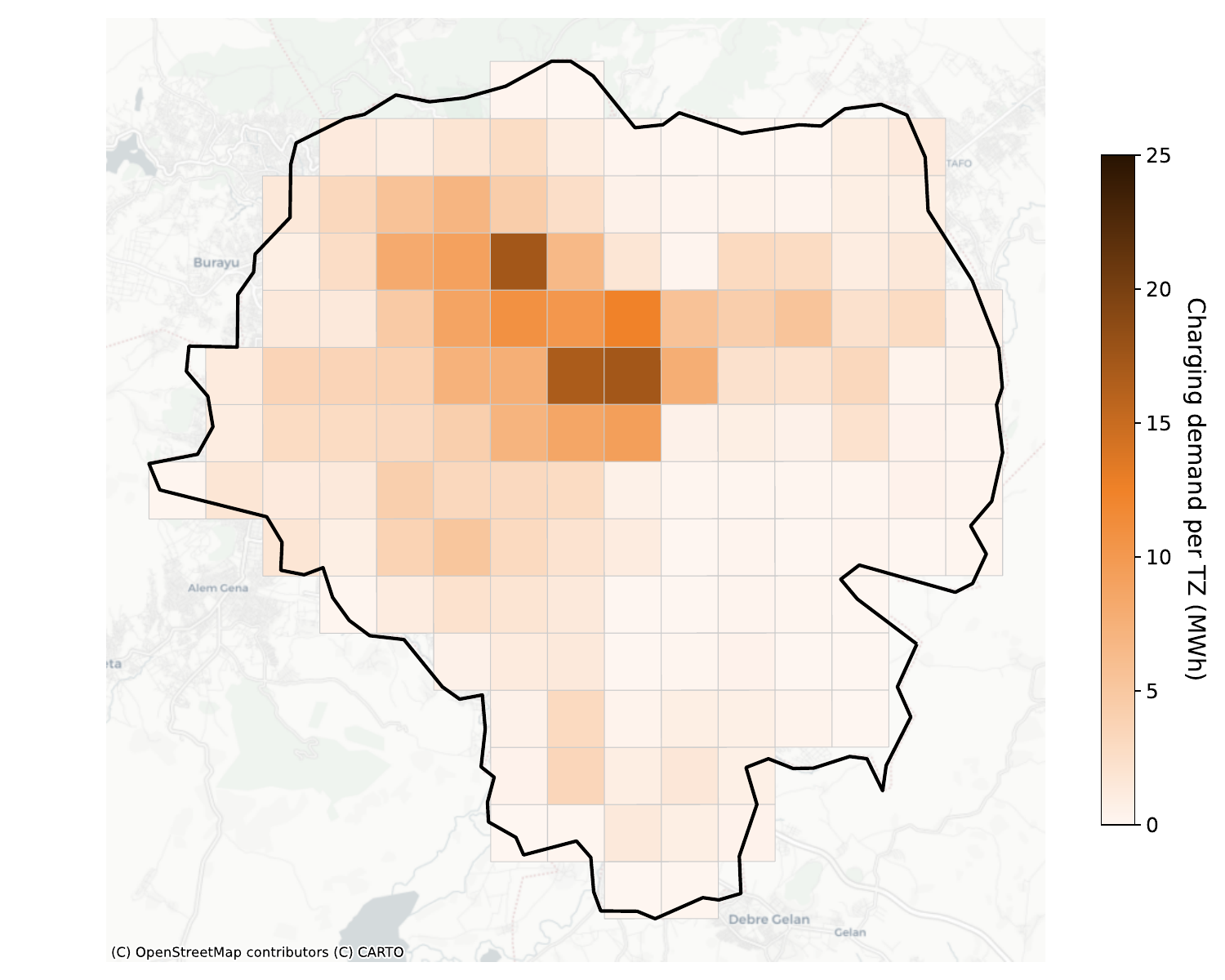}            
        \end{subfigure}
        \begin{subfigure}[t]{0.42\textwidth}
            \centering
            \caption{}
            \includegraphics[width=\textwidth]{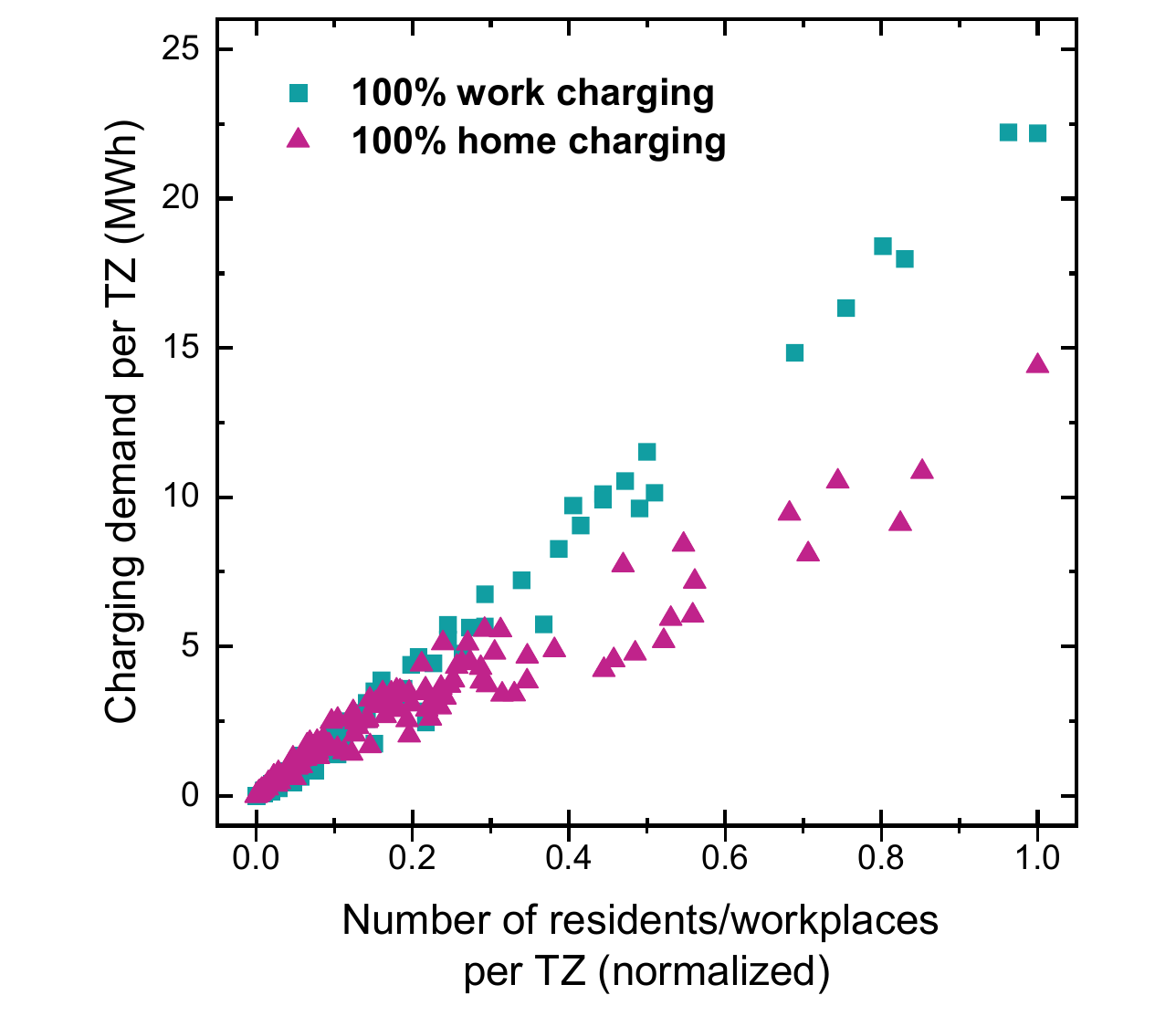}            
        \end{subfigure}
    \end{minipage}
    \caption{Map of the daily per-zone charging demand for the three charging scenarios: (a) 100\% home charging, (b) 100\% work charging, and (c) mixed charging. The traffic zone size is 1.95~km by 1.95~km. Panel (d) shows the per-zone charging demand as a function of the (normalized) number of residents and workplaces.}
    \label{fig:spatial_demand}
\end{figure}

Figure~\ref{fig:spatial_demand} presents the spatial distribution of daily charging demand for 100,000~EVs across the three charging scenarios, revealing variations in the magnitude and distribution pattern of the demand. The total daily charging demand is estimated at 353~MWh, driven by an average daily commuting distance of 17.4~km per vehicle (please refer to \ref{app:distance_dis} for the distribution of daily travel distances obtained from the mobility demand model), which aligns well with reported travel distances in other cities of comparable size to Addis Ababa \cite{Angel2016}. In the 100\% work charging scenario, demand is highly concentrated in the city center, where most workplaces are located, with a per-zone demand peaking at 22.2~MWh. In contrast, the 100\% home charging scenario shows a more dispersed pattern, with a maximum per-zone demand equal to 14.4~MWh. The mixed scenario presents an intermediate case, with a per-zone demand reaching 18.3~MWh. Notably, the spatial pattern in the mixed scenario closely resembles that of the 100\% workplace scenario.  This similarity arises because POIs are predominantly located near workplaces, leaving the 25\% share of home charging to only slightly disperse the charging demand across traffic zones.

Building on this spatial distribution, Fig.~\ref{fig:spatial_demand}.d also examines the per-zone charging demand as a function of the number of residents and workplaces for the 100\% home charging and 100\% workplace charging scenarios, respectively. In both cases, a positive correlation between the number of residents or workplaces and the charging demand per traffic zone is observed. However, the relationship does not follow a perfect linear trend. Notably, in the 100\% home charging scenario, as the population within a traffic zone increases, the growth rate of the per-zone charging demand slightly diminishes. This non-linearity can be attributed to the fact that while the number of EVs scales proportionally with the population, the per-vehicle charging demand is higher in less densely populated zones as these zones tend to be located farther from workplaces. In contrast, in the 100\% work charging scenario, the per-vehicle charging demand exhibits less sensitivity to the specific location of each traffic zone, primarily due to the more concentrated distribution of workplaces. This leads to a more linear trend between the number of workplaces and the charging demand per TZ. This linear trend is also driven by the underlying fact that, in Addis Ababa, the number of EVs scales linearly with the number of workplaces. These results underscore the importance of our spatially explicit approach to accurately capture the complex dependencies of the charging demand on the urban spatial structure.

\subsubsection{Temporal distribution}
\begin{figure}[htbp]
    \centering
    \includegraphics[width=0.5\textwidth]{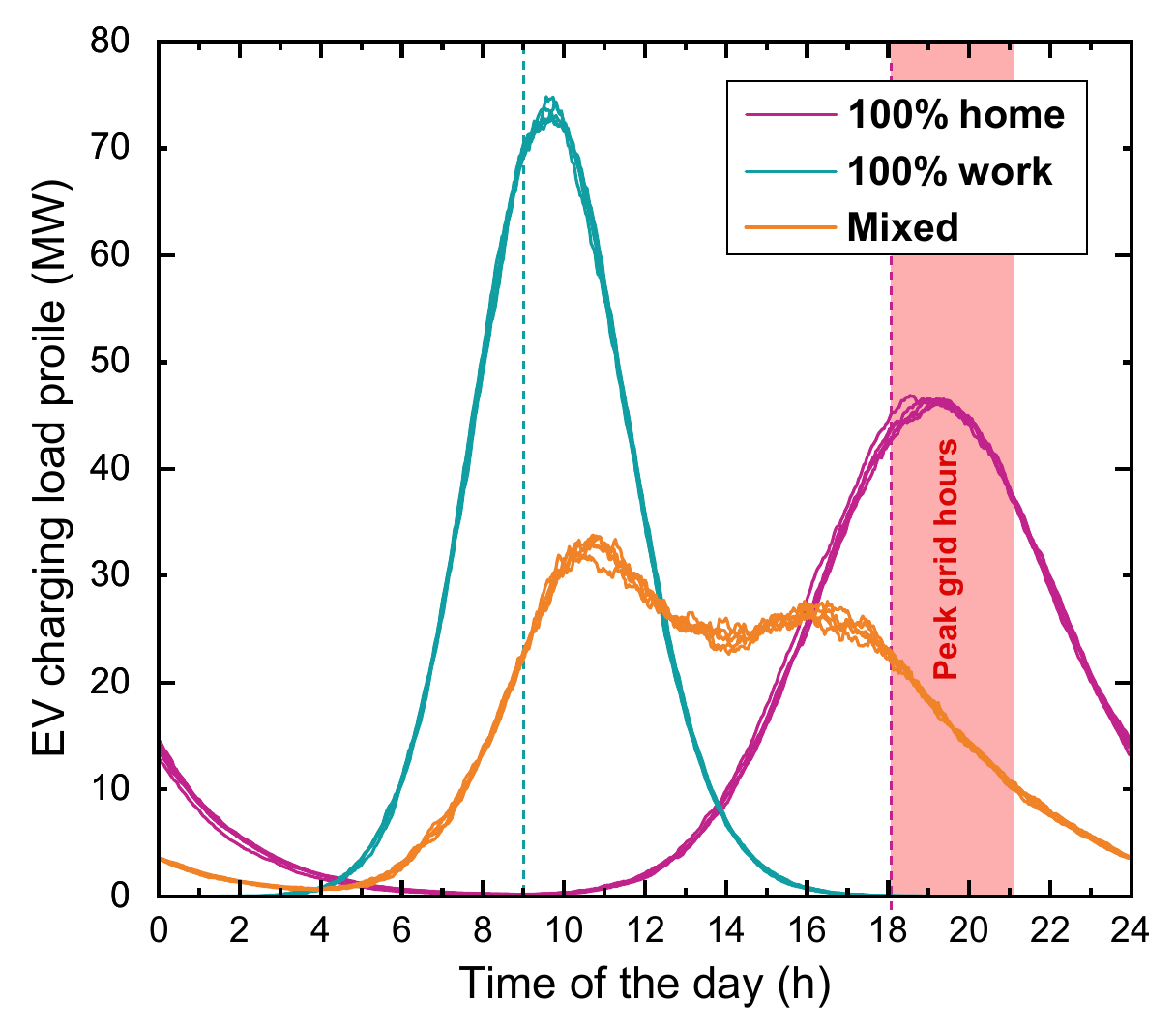}            
    \caption{EV charging load profiles for the three charging scenarios, aggregated across all TZs and electric vehicles. Each scenario is simulated five times to assess the impact of day-to-day variability. The average arrival times at work (blue dashed line) and at home (purple dashed line) are also shown alongside the typical peak grid hours in Addis Ababa.}
    \label{fig:temporal_demand}
\end{figure}

The charging scenario strongly shapes the EV charging load profile, influencing both the timing and intensity of demand peaks, as illustrated in Fig.~\ref{fig:temporal_demand}. Home charging induces a pronounced evening peak of 46.6±0.3~MW, coinciding with the typical peak grid hours. In contrast, workplace charging shifts this peak to the late morning, avoiding overlap with the evening grid peak. However, the peak intensity for workplace charging is higher, reaching 74.8±0.5~MW. This is due to the lower dispersion in arrival times and the higher charging power levels at workplaces compared to home charging. This suggests that a 100\% workplace charging strategy could also impose important stress on the grid, particularly under the assumption of uncontrolled charging adopted in this study. The mixed charging scenario presents a more balanced charging load profile. The charging demand is distributed more evenly throughout the day, with a peak in the morning reaching only 33.2±0.6~MW. This result underscores the potential benefits of charging a portion of the fleet at POIs, which creates a smoother demand profile and reduces the overlap of charging events with peak grid hours.

Since all EVs do not charge simultaneously, it is also interesting to note that the average peak demand per vehicle remains consistently lower than the average available charging power. Given that only about one-third of the 100,000~EVs charge daily, the peak demand for home charging is just 1.4~kW per charging EV, despite an average home charging power of approximately 6.0~kW. This value aligns with previous studies \cite{Pareschi2020} and results from the dispersion in vehicle arrival times and the relatively low per-vehicle charging demand, which limits the number of EVs charging at the same time. At workplaces, the peak demand per charging EV is higher (around 2.3~kW) but remains well below the average workplace charging power of approximately 12.8~kW.

\subsection{Implications on the required number of charging points}
The model we developed also provides guidance to estimate the charging infrastructure requirements, specifically the number of charging points needed. This estimate depends on the charging point-to-EV ratio, which varies based on expected usage patterns at the different charging locations. Table \ref{tab:charging_sockets} provides estimates of the charging point-to-EV ratios, as well as the total number of charging points per location under the three charging scenarios. 

\begin{table}[ht]
\centering
\setlength{\tabcolsep}{14pt} 
\caption{Charging points per EV and the total required number of charging points in three different charging scenarios at various locations. The values are derived from the average charging point-to-EV ratio obtained over 10 model runs.}
\begin{tabular}{lccc}
\hline
Charging location & Home & Work & POIs \\ \hline \hline
Charging point-to-EV ratio & 1.0 & 0.34 & 0.03 \\ \hline
Required charging points: \\ 
100\% home & 100,000 & - & - \\ 
100\% work & - & 32,000 & - \\ 
Mixed & 25,000 & 8,000 & 1,500 \\ \hline
\end{tabular}
\label{tab:charging_sockets}
\end{table}

In residential settings, it is likely that each electric vehicle will be equipped with a dedicated charging point, as privately owned charging infrastructure is rarely shared. While this arrangement offers convenience for EV users, it is inefficient from an infrastructure perspective. Hence, in the 100\% home charging scenario, such a setup would require 100,000 charging points. Each of these would likely need to be paired with either a dedicated charging station or, at a minimum, a reinforced charging socket. 

At workplaces, the charging point-to-EV ratio is approximately three times lower compared to residential settings. This benefit stems from the fact that only a fraction of the EV fleet requires charging on any given day, as calculated by our charging-decision model, which serves as a basis for estimating the minimum charging point-to-EV ratio. Additionally, since multiple charging points may be integrated into a single charging station at workplaces, this further contributes to a decrease in the overall number of required charging stations. It is important to highlight that the charging point-to-EV ratio is influenced by both daily charging demand and the battery capacities of the vehicles, making it highly context-dependent. In the case of Addis Ababa, if battery capacities continue to increase in line with global trends, the current estimated ratio is expected to decline further.

At POIs, the estimated charging point-to-EV ratio is ten times lower than at workplaces. Assuming that users release their charging stations immediately upon completing their charging session—which is most likely to occur at public charging locations—the charging point-to-EV ratio is determined by the minimum number of EVs charging simultaneously, as estimated by our model. Notably, this number is particularly low for POIs due to a combination of high charging powers and the dispersed distribution of arrival times. As a result, POIs emerge as potentially the most efficient locations for minimizing the required number of charging stations.

\subsection{Potential for PV-based EV charging}
\begin{figure}[htbp]
    \centering
    \includegraphics[width=0.65\textwidth]{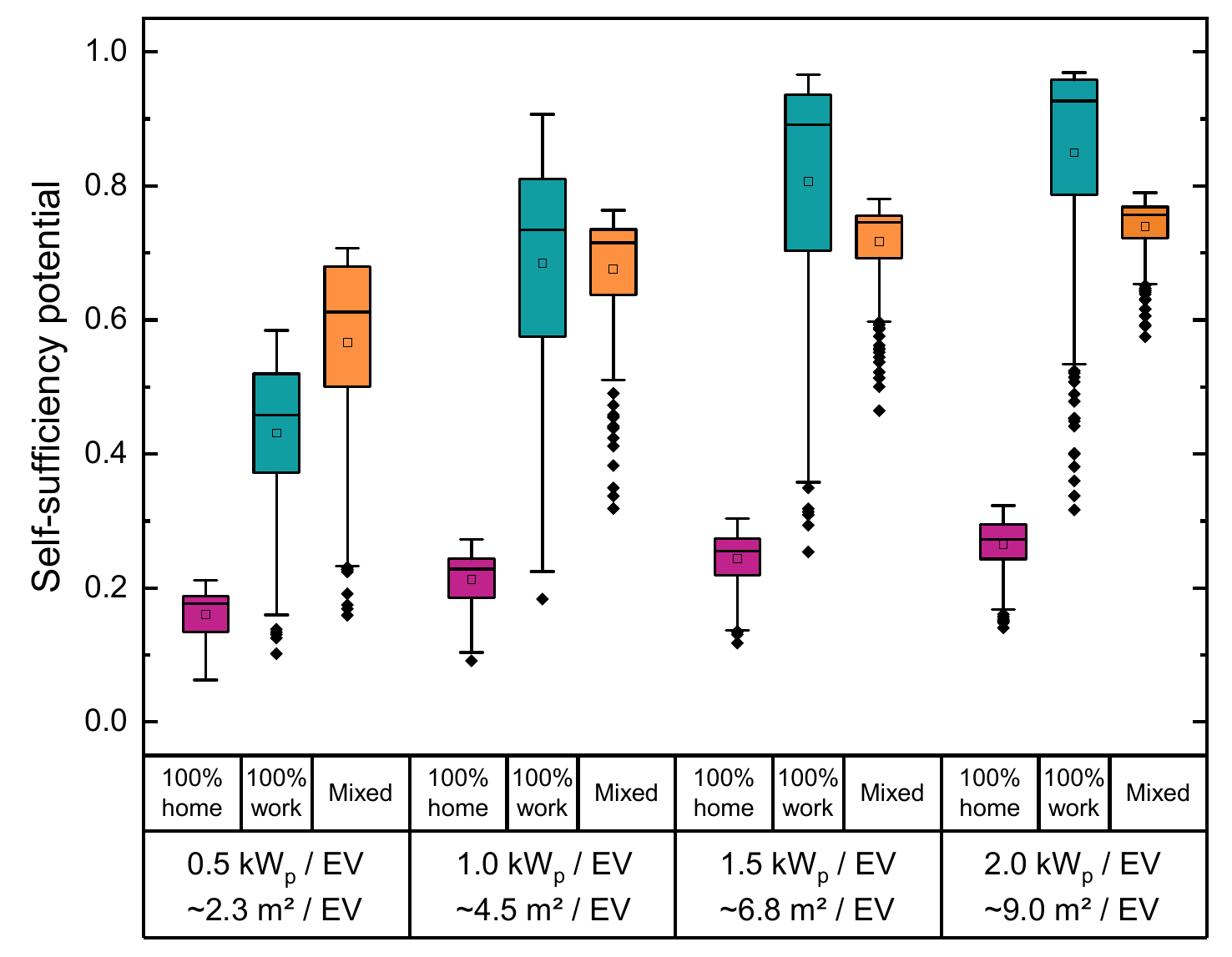}            
    \caption{Self-sufficiency potential as a function of the installed nominal PV capacity per EV for the different charging scenarios. Boxplots are based on the daily self-sufficiency potential for all 262~weekdays of the year 2020. The boxplots represent the daily self-sufficiency potential for all 262~weekdays of 2020. They show the averages (circle markers), medians (horizontal lines), interquartile ranges (IQR, box outlines), and outliers (black dots). The whiskers extend from the hinge to the highest and lowest values within 1.5 × IQR. Nominal PV capacities range from 0.5 to 2.0~kW$_p$/EV, corresponding to surface areas of approximately 2.3 to 9.0~m$^2$/EV.}
    \label{fig:self_sufficiency}
\end{figure}

Leveraging PV energy for EV charging in regions with abundant solar resources, such as Addis Ababa, offers a promising opportunity to reduce the grid electricity demand for charging while aligning with broader goals of decarbonization. The self-sufficiency potential, as defined in Section \ref{methodology}, is depicted in Fig.~\ref{fig:self_sufficiency}, which presents a box plot of the self-sufficiency potential calculated for all weekdays over a year. The analysis considers the three charging scenarios and various levels of PV capacity, ranging from 0.5~kW$_p$/EV to 2.0~kW$_p$/EV, approximately equivalent to one to four modern silicon solar panels per EV. In Addis Ababa, this range produces between 2.3~kWh (0.5~kW$_p$/EV) and 9.1~kWh (2.0~kW$_p$/EV) of PV energy per EV per day on average.

Despite daily and scenario-specific variations, a key finding is that an important share of the charging demand can reliably be met by PV energy. Even in the 100\% home charging scenario (where temporal complementarity between PV production and EV charging is poor due to predominantly nighttime charging), the average self-sufficiency potential ranges from 16.0\% (0.5~kW$_p$/EV) to 26.5\% (2.0~kW$_p$/EV). The 100\% work charging scenario demonstrates substantially higher potential, with averages ranging from 43.1\% to 84.9\%. The mixed charging scenario achieves intermediate levels, with an average self-sufficiency potential ranging from 56.6\% to 73.9\%.

Notably, in the 100\% work charging scenario, self-sufficiency levels exceeding 95\% are achieved on several days for PV capacities of 1.0~kW$_p$/EV or higher, owing to the strong temporal alignment between PV generation and EV charging needs. Conversely, the mixed charging scenario includes some evening and nighttime charging, reducing its temporal alignment with PV production. However, this scenario exhibits less day-to-day variability, as the broad charging load profile ensures that part of the available solar energy is always used for EV charging.

\begin{figure}[htbp!]
    \centering
    \begin{minipage}[b]{1.0\textwidth} 
        \centering
        \begin{subfigure}[t]{0.52\textwidth}
            \centering
            \caption{}
            \includegraphics[width=\textwidth]{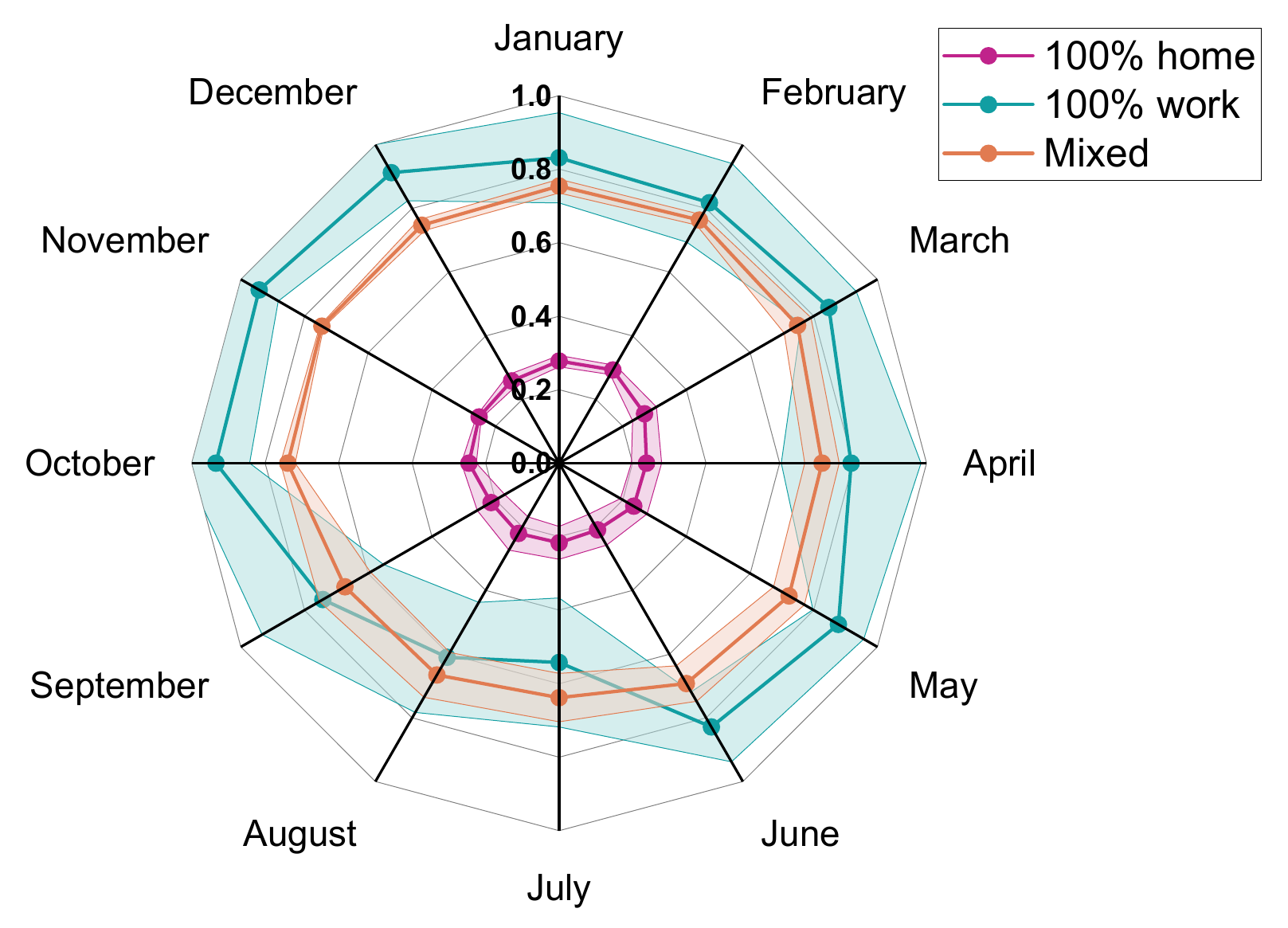}            
        \end{subfigure}
        \begin{subfigure}[t]{0.47\textwidth}
            \centering
            \caption{}
            \includegraphics[width=\textwidth]{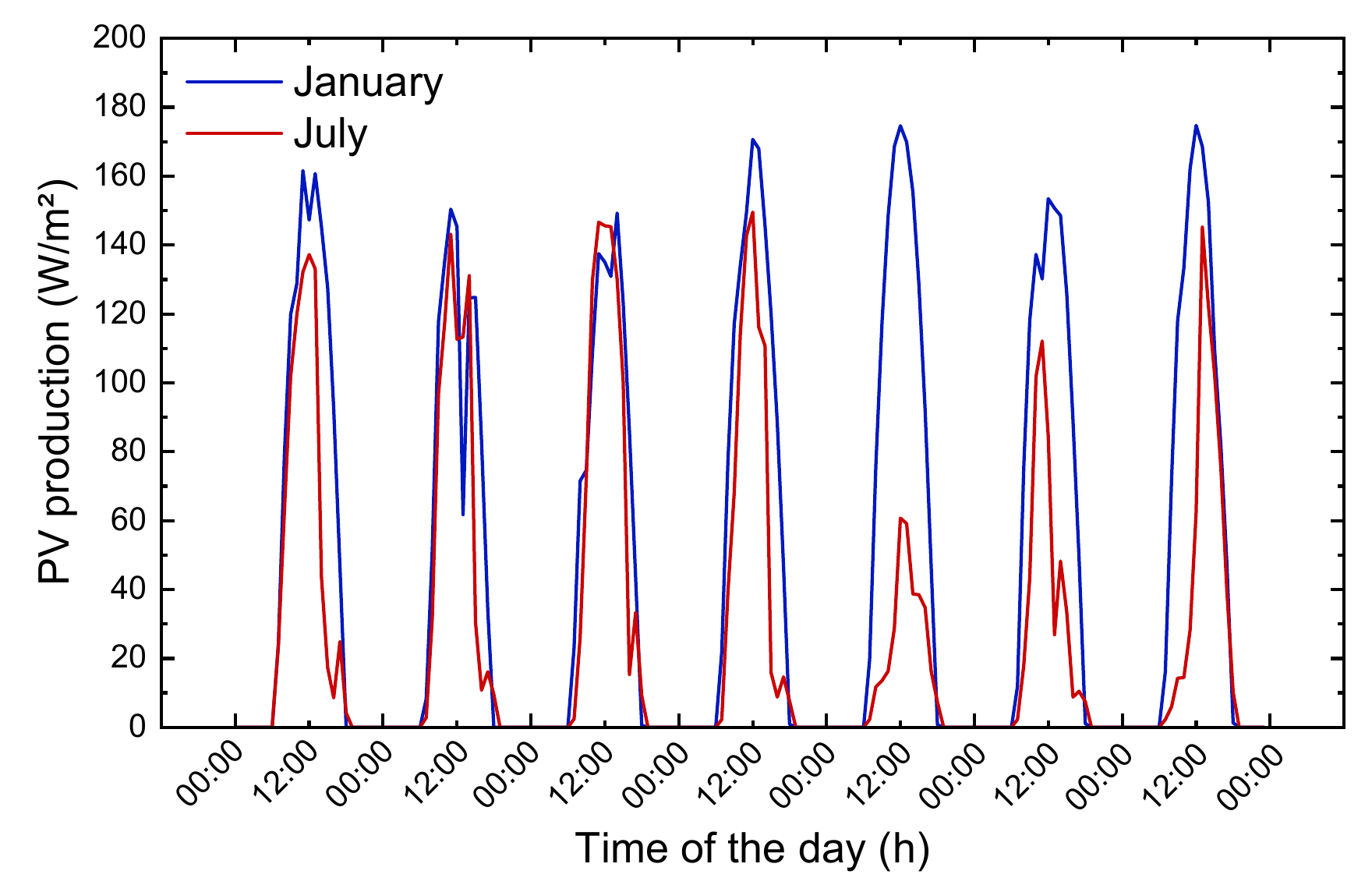}            
        \end{subfigure}
    \end{minipage}
    \caption{Influence of monthly variations on the self-sufficiency potential: (a) radar plot of the monthly self-sufficiency potential (average and standard deviation) for the three charging scenarios, based on a PV capacity of 1.5~kW$_p$/EV ; (b) Hourly PV production profiles for a representative week in January and July. }
    \label{fig:self_sufficiency_monthly}
\end{figure}

A closer examination of monthly variations further highlights the differences across charging scenarios, as illustrated in Fig.~\ref{fig:self_sufficiency_monthly}.a. As anticipated, the 100\% work charging scenario achieves higher average self-sufficiency levels compared to the mixed charging scenario in most months. However, it also exhibits greater variability, with larger day-to-day variations and more pronounced month-to-month fluctuations in the average values. Specifically, the average self-sufficiency in the 100\% work charging scenario ranges from 54\% to 94\%, with the lowest levels occurring during July and August, coinciding with the rainy season. In contrast, the mixed charging scenario offers a more stable self-sufficiency range of 64\% to 76\%. The attenuated complementarity between PV generation and work charging during the rainy season can be attributed to limited PV production on some mornings. This seasonal mismatch is further illustrated in the weekly PV production curves for January and July (Fig.~\ref{fig:self_sufficiency_monthly}.b), which underscore the advantages of charging at POIs to mitigate the effects of weather variability during the rainy season. 

\subsection{Discussion}
The results indicate that the introduction of 100,000~EVs in Addis Ababa would lead to a daily charging demand of 353~MWh. Relative to the city's current electricity demand profile, this corresponds to an approximate 1.5\% rise in the overall daily electricity demand. While this increase appears manageable and suggests that the deployment of the first wave of EVs could proceed without major challenges, it nonetheless requires proper planning, as the additional demand is not negligible. Addis Ababa's electricity grid already operates near capacity, as evidenced by the frequent power interruptions, and the spatial analysis shows that the concentrated charging demand could further strain specific parts of the grid. Furthermore, this 1.5\% increase does not account for the electrification of public transport, nor does it consider the potential for private vehicle ownership to exceed 100,000~EVs, which would amplify the charging demand. This analysis provides a foundation for supporting future EV adoption by offering scalable estimates and identifying high-demand charging locations.

The charging location plays an important role in shaping the temporal distribution of the charging demand. Home charging is likely to coincide with the existing evening peak hours, while workplace charging could introduce a new peak in the morning, contrasting with the current electricity demand profile (see Fig.~\ref{fig:aa_load_profile}). It is important to emphasize that the EV charging load profiles presented here represent the most probable scenario. However, statistical fluctuations in arrival times could lead to higher peak demands on certain days. In the worst-case scenario (where all vehicles charging on a given day charge simultaneously) peak demand could increase by approximately 193~MW in the 100\% home charging scenario and 410~MW in the 100\% work charging scenario, a 31.3\% increase in the current peak demand. Although such scenarios are unlikely, they underscore the need for a robust electricity grid to accommodate potential extreme cases. Demand-side strategies such as implementing smart charging systems or promoting the use of public charging stations could also mitigate the risk of extreme peak demands while reducing the need for costly grid upgrades. Notably, existing evidence suggests that when public charging stations are available they are likely to be used even when other charging facilities are accessible \cite{Liu2022}.

\begin{figure}[htbp]
    \centering
    \includegraphics[width=0.55\textwidth]{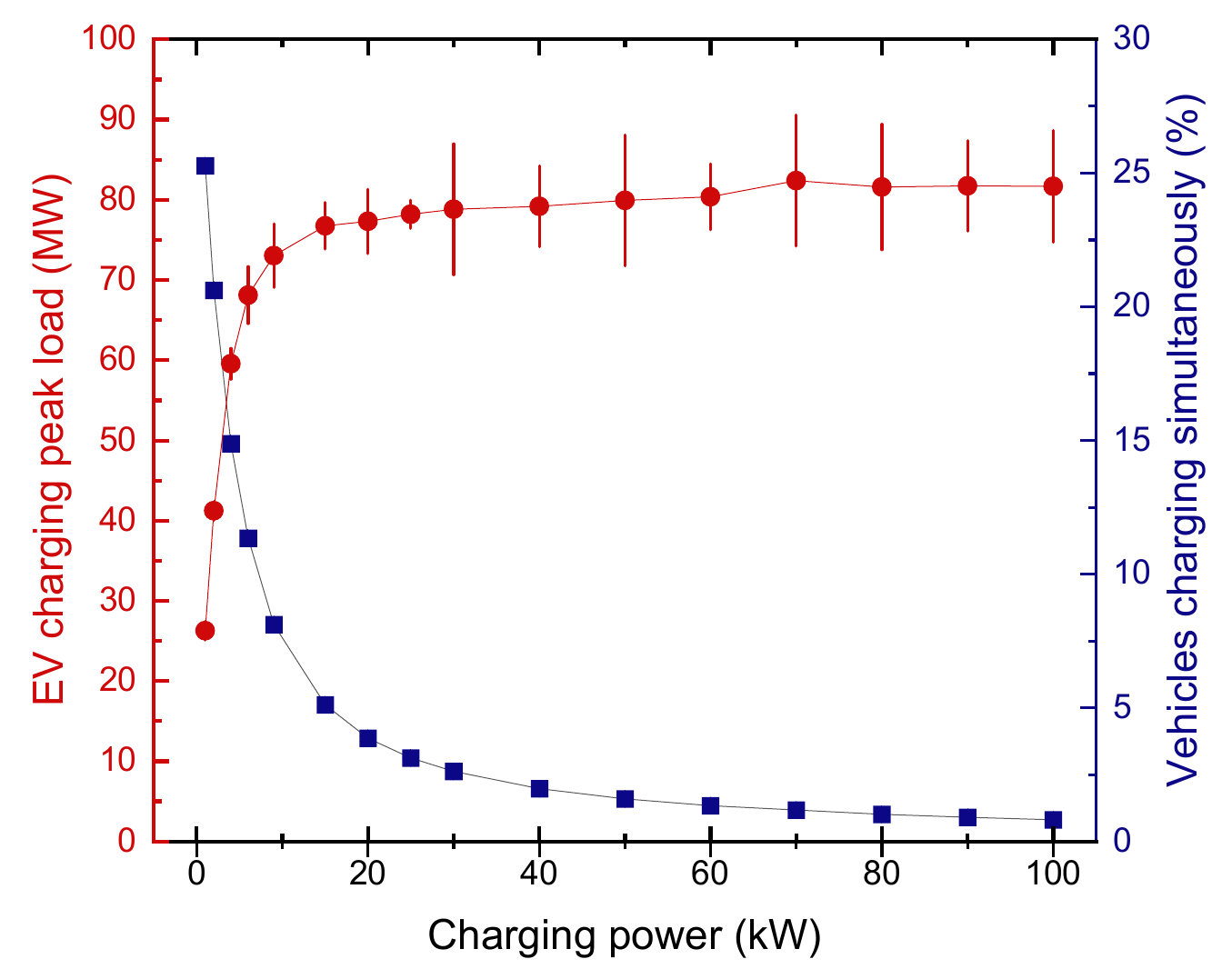}            
    \caption{Dependence of EV charging peak load (red, left axis) and the percentage of EVs charging simultaneously (blue, right axis) on the charging power. Error bars represent daily variability in EV load profiles. Results are derived from a dummy scenario with a single charging power level and a tight distribution in arrival times (standard deviation equal to 1.8~hours), representing a highly sensitive case.}
    \label{fig:peak_charging_load}
\end{figure}

Charging at POIs could also substantially reduce the number of required charging stations. This reduction arises not only from the assumed ideal charging behavior but also from the greater share of fast chargers typically installed at POIs. However, the integration of fast chargers introduces a potential trade-off: while reducing the number of charging points improves cost efficiency and profitability for charge point operators by maximizing energy throughput per station, it may also lead to an increase in peak power demand. Interestingly, as shown in Fig.~\ref{fig:peak_charging_load}, peak power demand rises with a charging power only up to approximately 20~kW, beyond which it stabilizes and becomes relatively independent of further increases in charging power. This behavior is driven by the interplay between charging power and the maximum number of EVs charging simultaneously, the latter decreasing as charging power increases. This finding suggests that deploying fast chargers at locations requiring high charging powers, such as POIs, may not necessarily exacerbate peak power demand, challenging the assumption that high-power charging strains the grid. Instead, fast chargers can optimize charging station usage while maintaining a manageable peak load.

PV-based charging has significant potential to meet a large share of the EV charging demand, particularly at workplaces and POIs. However, our analysis indicates that beyond a PV capacity of 1.5~kW$_p$/EV, additional PV capacity yields only marginal gains in self-sufficiency. This limitation arises from the inherent mismatch between PV generation and EV charging demand, which persists across our three archetypal scenarios. Indeed, on average, daily solar energy generation exceeds daily EV charging demand at 1.0~kW$_p$/EV and above. Nonetheless, more advanced charging scenarios or strategies could improve PV self-sufficiency. In particular, smart charging approaches such as dynamic scheduling and demand-responsive charging could help shift charging times to better coincide with peak solar production hours.


\section{Conclusion} 
\label{conclusion}
Transport electrification is gaining momentum in Africa, offering a key opportunity to cut greenhouse gas emissions and reduce dependence on imported fossil fuels when paired with renewable energy. However, many African regions lack comprehensive transport and energy data, posing a significant barrier to electric mobility planning. This article introduces a novel modeling framework specifically designed for regions with limited data availability. The framework enables the simulation of mobility patterns, spatio-temporal charging needs, and the complementarity between EVs and PV energy by leveraging open-source geospatial data. Additionally, it facilitates scenario-based analyses, providing actionable insights for policymakers and energy planners to guide energy infrastructure development.

The framework is applied to Addis Ababa, simulating 100,000~EVs and demonstrating how one of Africa's largest cities could meet a substantial share of its private passenger transport energy demand with locally generated PV energy. The analysis considers three distinct charging scenarios: home charging, workplace charging, and a mixed scenario where 50\% of charging occurs at POIs. Based on the results, the following key findings can be drawn:
\begin{enumerate}
    \item While the per-vehicle charging demand is relatively low (3.53~kWh) due to short daily commuting distances, the total demand for 100,000~EVs results in a 1.5\% increase in overall electricity consumption.
    \item The charging scenario significantly affects the spatial charging demand distribution. Workplace and POI charging concentrates the charging demand in the city center, whereas home charging shifts it to residential areas and disperses slightly the demand. 
    \item The charging scenarios also shapes the charging load profile. Home charging leads to an evening peak aligned with current electricity peak hours, while workplace charging generates an earlier but higher peak. Charging at POIs leads to a more evenly distributed load, highlighting the benefit of charging at public places.
    \item The number of required charging points varies by location. Workplace charging requires about one charger per three EVs, unlike home charging, where each EV has a dedicated unit. At POIs, fast chargers and dispersed arrivals further reduce charging points by a factor of ten.
    \item Workplace and POI charging align well with PV production. The average daily self-sufficiency potential exceeds 80\% at workplaces and 70\% in the mixed charging scenario for PV capacities of 1.5~kW$_p$/EV or above. While the mixed charging scenario exhibits slightly lower self-sufficiency, it shows less variability, particularly during the rainy season.
\end{enumerate}

We acknowledge several limitations in our study. While the results for Addis Ababa were compared with existing literature where possible, some were derived using proxy data (e.g., charging power distribution) or informed assumptions (e.g., average arrival times at home and work). One limitation of this study is the lack of real-world data specific to Addis Ababa, particularly concerning mobility patterns, which would allow for a more precise calibration of the model. Should more data become available, the model could be further refined to incorporate local specificities, such as the detailed distribution of arrival times or the characteristics of the EV fleet. The model could also benefit from more granular geospatial inputs, which would enable the allocation of EVs to traffic zones based on socio-economic factors. Regarding the assessment of EV–PV complementarity, our analysis is limited to the year 2020. Extending the study to assess multiple years would provide a more comprehensive evaluation. However, this limitation is unlikely to significantly impact the results, given that annual solar irradiance variations are minimal (data from PVGIS-SARAH3 indicate fluctuations of about 2\% per year in Addis Ababa).

For future research, our model could be applied to other African cities with diverse vehicle types, such as motorbikes, or different urban structures, including those with multiple business districts. Expanding the model to capture a broader scope of mobility demand would also be valuable, provided that relevant data becomes available. For instance, considering mobility associated with shopping or entertainment is expected to slightly increase the daily travel distance, as these activities often involve additional trips beyond commuting. This also includes accounting for weekend mobility demand and charging behavior, as weekend mobility patterns differ significantly from those on weekdays. Notably, home charging during weekends could better align with residential PV generation, enhancing the self-sufficiency of this charging location. Additionally, considering alternative weekday mobility patterns, such as commuting to park-and-ride facilities, could offer valuable insights into the long-term development pathways of the overall passenger transport system, integrating both public and private mobility. Further analysis could also explore the impact of smart charging strategies on charging load profiles and EV-PV complementarity, providing insights into mitigating grid strain while maximizing solar energy use. Expanding the model to individual charging stations with techno-economic analyses could further refine these insights, offering more detailed and practical guidance for PV-based charging stations. In this context, pairing PV with stationary batteries could also be a cost-effective strategy to investigate, especially given the rapidly declining costs of battery storage.


\appendix

\setcounter{figure}{0}
\renewcommand{\thesection}{Appendix \Alph{section}} 

\section{Charging probability function}
\label{app:charging_probability}

\begin{figure}[H]
    \centering
    \includegraphics[width=0.6\textwidth]{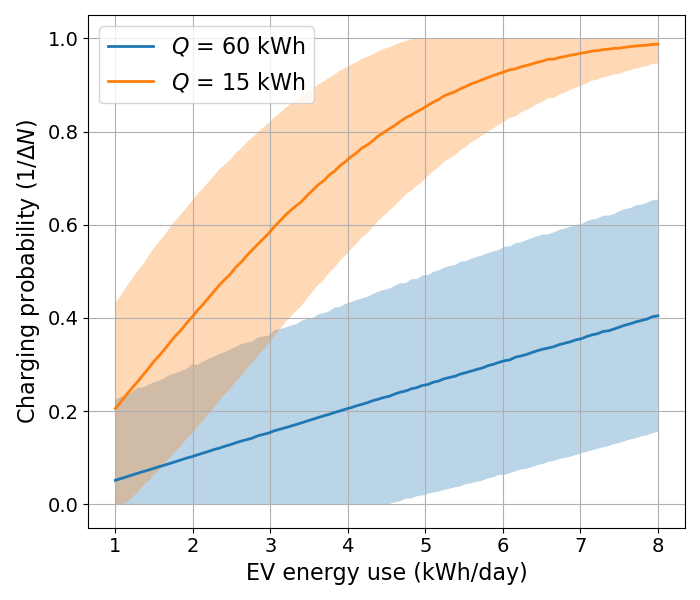}           
    \caption{Charging probability function as a function of daily EV energy use, for two battery capacities. The plot displays the average charging probability (solid line) along with the standard deviation. Results are derived from random sampling of 50,000 values.}
    \label{fig:charging_probability}
\end{figure}

\section{EV fleet share dynamics}
\label{app:ev_fleet_share}

Considering a fixed stock of vehicles (electric and non-electric), the evolution of the EV fleet share can be modeled considering the balance between the vehicle fleet renewal rate (assumed to occur at the same rate across all vehicle types) and the introduction of new EVs

\begin{equation}
    \frac{ds}{dt} = \lambda (\sigma - s(t))
\end{equation}
where $s(t)$ represents the EV share in the fleet at time $t$, $\lambda$ is the fleet renewal rate (the average share of vehicles replaced per unit time), and $\sigma$ denotes the EV share in new vehicle registrations.

Solving this equation with the initial condition $s(0)=s_0$ gives:
\begin{equation}
    s(t) = \sigma + (s_0-\sigma)e^{-\lambda t}
\end{equation}
which describes a growing function converging towards the EV share in new vehicle registrations $\sigma$, with a characteristic time scale of $1/\lambda$.

The numerical results, assuming an average vehicle lifetime of 20~years ($\lambda=1/20$) and $s_0=0$, indicate that the transition towards an EV-dominated fleet is strongly influenced by the EV share in new registrations ($\sigma$) and, consequently, the effectiveness of EV adoption policies. If all newly registered vehicles are EVs ($\sigma=1$), reaching 100,000~EVs (approximately 1/6 of the current vehicle stock in Addis Ababa) would take just 3.6 years. For $\sigma=0.5$, the time required is around 8.1~years. However, for $\sigma=0.2$, the time extends significantly to 35.8~years.

\section{OSM Query for workplace and POI extraction}
\label{app:osm}

\subsection*{Workplaces}
\begin{small}
\begin{verbatim}
tags = {
    "building": ["industrial", "office"],
    "company": True,
    "landuse": ["industrial", "commercial", "retail"],
    "industrial": True,
    "office": True,
    "amenity": [
        "university", "research_institute",     
        "hospital", "townhall",               
        "conference_centre", "factory",                
        "corporate_office", "government",             
        "bank", "police",                 
        "fire_station", "post_office",            
        "call_centre", "logistics_centre"        
    ]
}
\end{verbatim}
\end{small}

\subsection*{Points of interest}
\begin{small}
\begin{verbatim}
tags = {
    "amenity": [
        "fuel", "parking", 
        "parking_entrance", "bicycle_parking",        
        "college", "university", 
        "school", "kindergarten", 
        "library", "music_school", 
        "language_school", "clinic", 
        "dentist", "doctors", 
        "hospital", "pharmacy", 
        "veterinary",
        "cafe", "ice_cream", 
        "internet_cafe", "restaurant", 
        "fast_food", "bar", 
        "pub", "biergarten",
        "theatre", "cinema", 
        "music_venue", "nightclub", 
        "casino", "gambling", 
        "stripclub", "arts_centre", 
        "community_centre", "social_centre", 
        "exhibition_centre", "attraction", 
        "viewpoint", "aquarium", 
        "beach_resort", "gallery", 
        "museum", "theme_park", 
        "zoo", "artwork"
    ],
    "shop": [
        "supermarket", "mall", 
        "department_store", "convenience"
    ],
    "tourism": [
        "hotel", "guest_house", 
        "hostel", "motel", 
        "camp_site", "apartment"
    ],
    "leisure": [
        "stadium", "sports_centre", 
        "swimming_pool", "fitness_centre"
    ]
}
\end{verbatim}
\end{small}

\section{Daily commuting distance distribution}
\label{app:distance_dis}

\begin{figure}[H]
    \centering
    \includegraphics[width=0.6\textwidth]{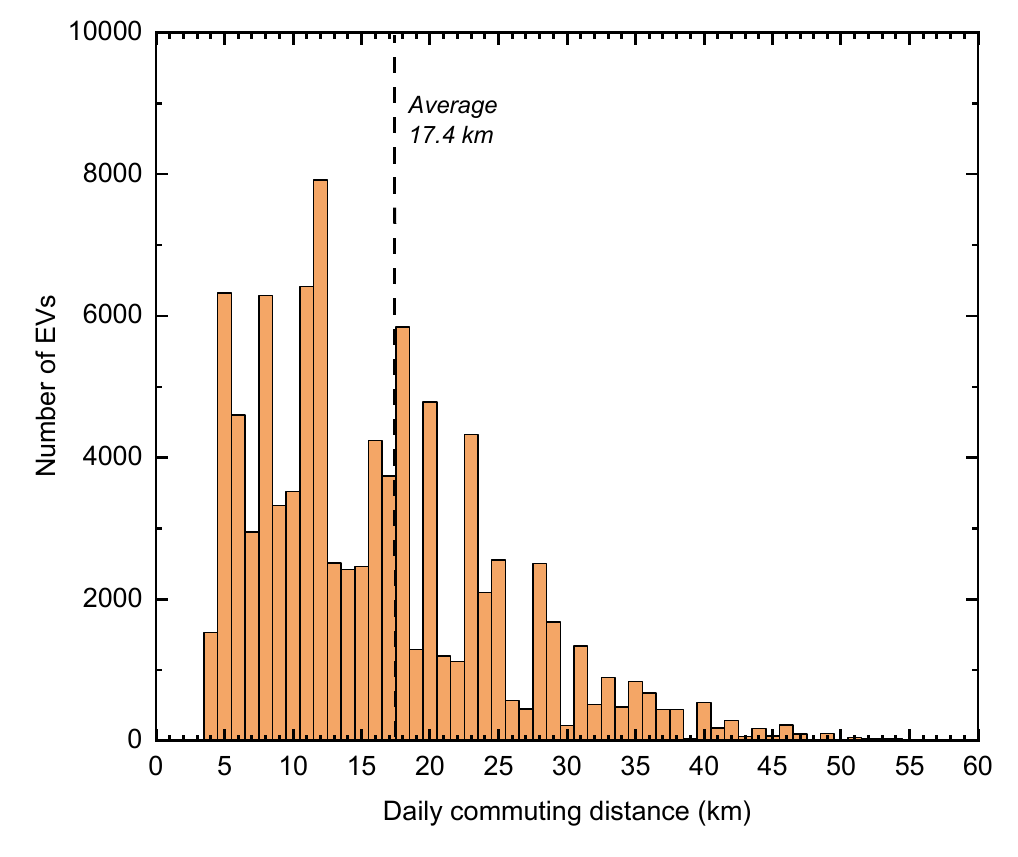}           
    \caption{Histogram of the daily commuting distance (two-way) for the 100,000 simulated EVs. }
    \label{fig:commuting_distance_distribution}
\end{figure}

\section{Electricity demand profile for Addis Ababa}
\label{app:demand_profile}

Hosting 30\% of the urban population of Ethiopia, Addis Ababa, the capital of Ethiopia and the country's political and economic centre, the seat of head offices of African Union \& UN Economic commission for Africa, home to several small to high-scale industries, is one of the fastest growing cities on the continent. Its geographic location in the center of Ethiopia, combined with lack of development in other urban centers have given the capital the majority of social and economic infrastructure in the country. This makes Addis Ababa the largest electric load center in the country. 

Due to the absence of a specific load curve for the study area (Addis Ababa administrative region), the national load curve for 2024, which has a peak demand of 4,560~MW, was scaled to match the peak demand of Addis Ababa. The peak demand for Addis Ababa was estimated based on the extended Addis Ababa region's 2024 peak demand of 2,100~MW \cite{JapanInternationalCooperationAgency2017}, adjusted proportionally to the population within the study area. The study area comprises 5.54~million people, while a GIS analysis indicates that the extended region encompasses 8.88~million inhabitants. As a result, the study area represents 28.7\% of the national demand, with an estimated peak intensity of 1,310~MW and a total daily energy demand of 23,198~MWh.

\section*{CRediT authorship contribution statement}
\label{authorship}
\textbf{Jérémy Dumoulin:} Conceptualization, Methodology, Software, Data Curation, Writing - Original Draft, Visualization. \textbf{Dawit Habtu: }Conceptualization, Writing - Original Draft, Investigation. \textbf{Kanchwodia Gashaw: }Writing - Review \& Editing, Investigation. \textbf{Noémie Jeannin:} Writing - Review \& Editing. \textbf{Alejandro Pena-Bello:} Writing - Review \& Editing. \textbf{Christophe Ballif:} Writing - Review \& Editing, Funding acquisition. \textbf{Nicolas Wyrsch:} Writing - Review \& Editing, Supervision, Funding acquisition. \textbf{Ingeborg Graabak: } Conceptualization, Writing - Original Draft.

\section*{Declaration of competing interest}
\label{competing-interest}
The authors declare no competing interests.

\section*{Acknowledgments}
\label{acknowledgments}
This work was supported by the HORIZON OpenMod4Africa project (Grant number 101118123), with funding from the European Union and the State Secretariat for Education, Research and Innovation (SERI) for the Swiss partners. The authors would like to thank Dr. Maxime Lenormand for the insightful discussions on the self-calibrated gravity model. Our gratitude is also extended to Dr. Giulia Vaglietti and the Fondazione Eni Enrico Mattei (FEEM) for their valuable support during the writing process. 

\section*{Data availability}
\label{data}
The current version of the open-source modeling framework is publicly accessible on GitHub (\url{https://github.com/jeremydumoulin/evpv-simulator}). Additionally, the specific version of the code, along with all data used in this work is available on Zenodo (\url{https://doi.org/10.5281/zenodo.14826594}), ensuring the complete reproducibility of our analyses.


\bibliographystyle{elsarticle-num} 
\bibliography{references}

\end{document}